%% file: sigir2024-resource-selection-beir-dataset.tex
\documentclass[acmlarge]{acmart}
\usepackage{tabularx}
\usepackage{subfigure}
\usepackage{multirow}
\usepackage{enumitem}
\usepackage{pifont}
\usepackage{multirow}
\usepackage{array}
\usepackage{enumitem}
\usepackage{nameref}
\usepackage{adjustbox}
\usepackage{graphicx}
\usepackage{xcolor}
\usepackage{tikz}
\usetikzlibrary{positioning}

\newcommand{\cmark}{\ding{51}}
\newcommand{\xmark}{\ding{55}}
\newcommand{\beir}{{\texttt{BEIR}} }

\newcolumntype{P}[1]{>{\raggedright\arraybackslash}p{#1}}
\newcolumntype{M}[1]{>{\raggedright\arraybackslash\ttfamily}m{#1}}

\AtBeginDocument{%
  }

\hypersetup{draft}
\begin{document}

\title{FeB4RAG: Evaluating Federated Search in the Context of Retrieval Augmented Generation}

\author{Shuai Wang}
\affiliation{
	\institution{The University of Queensland}
	\streetaddress{4072 St Lucia}
	\city{Brisbane}
	\state{QLD}
	\country{Australia}}
\email{shuai.wang@uq.edu.au}

\author{Ekaterina Khramtsova}
\affiliation{
	\institution{The University of Queensland}
	\streetaddress{4072 St Lucia}
	\city{Brisbane}
	\state{QLD}
	\country{Australia}}
\email{e.khramtsova@uq.edu.au}

\author{Shengyao Zhuang}
\affiliation{
	\institution{CSIRO}
	\streetaddress{4072 St Lucia}
	\city{Brisbane}
	\state{QLD}
	\country{Australia}}
\email{shengyao.zhuang@csiro.au}

\author{Guido Zuccon}
\affiliation{
	\institution{The University of Queensland}
	\streetaddress{4072 St Lucia}
	\city{Brisbane}
	\state{QLD}
	\country{Australia}}
\email{g.zuccon@uq.edu.au}


\begin{abstract}
	
Federated search systems aggregate results from multiple search engines, selecting appropriate sources to enhance result quality and align with user intent. With the increasing uptake of Retrieval-Augmented Generation (RAG) pipelines, federated search can play a pivotal role in sourcing relevant information across heterogeneous data sources to generate informed responses. However, existing datasets, such as those developed in the past TREC FedWeb tracks, predate the RAG paradigm shift and lack representation of modern information retrieval challenges. 

To bridge this gap, we present FeB4RAG~\footnote{\url{https://github.com/ielab/FeB4RAG}}, a novel dataset specifically designed for federated search within RAG frameworks. This dataset, derived from 16 sub-collections of the widely used \beir benchmarking collection, includes 790 information requests (akin to conversational queries) tailored for chatbot applications, along with top results returned by each resource and associated LLM-derived relevance judgements . 
Additionally, to support the need for this collection, we demonstrate the impact on response generation of a high quality federated search system for RAG compared to a naive approach to federated search. We do so by comparing answers generated through the RAG pipeline through a qualitative side-by-side comparison. Our collection fosters and supports the development and evaluation of new federated search methods, especially in the context of RAG pipelines.


\end{abstract}



\maketitle

\input{sections/introduction.tex}

\input{sections/methods.tex}
\input{sections/experimental_setup.tex}

\input{sections/collection_use.tex}

\input{sections/results.tex}
\input{sections/related-works.tex}
\input{sections/conclusion.tex}


\bibliographystyle{ACM-Reference-Format}
\bibliography{sigir2024-resource-selection-beir-dataset}

\end{document}

%% file: sections/introduction.tex
\section{Introduction}
\label{sec:demo}
Often information is spread across multiple information systems and repositories~\cite{bhavnani2009information}. For example consider a university (and example familiar to most readers) and the need for the stakeholders of that university to access internal information: Academic staff wanting to access information about policies and procedures, but also about research grant opportunities, and library catalogues and so forth; Professional support staff wanting to access data repositories associated to HR and Finance functions, along again with policies and procedures; Students needing to find and access information about courses, scholarships, fee payments, events, etc. In a typical organisation of the size of a university, this information does not reside on a unique information system: instead it is spread across several resources: from internal websites (some publicly facing too), to library catalogues, e-learning systems, shared cloud space solutions (e.g., Microsoft's Sharepoint and Teams), databases, among many other data management solutions. While this problem of\textit{ information scattering}~\cite{bhavnani2009information} is even more exacerbated in larger organisations, we are increasingly witnessing information scattering affecting also individuals with their personal files and information spread across multiple information services (multiple email accounts, cloud storage providers, multimedia management applications, etc.). 

Federated search attempts to address information scattering by providing a search engine technology that is able to gather information from multiple resources, and assemble this information into a unique set of search results. Federated search has had extensive attention in the early 2000s, where the focus was to create meta Web search engines, that federated separate search engines available on the Web~\cite{shokouhi2011federated,demeester2013overview,demeester2014overview}; but this attention has largely lessened since. We believe that the recent, rapid increasing development and uptake of Retrieval Augmented Generation (RAG) pipelines, especially in the context of language-based information agents such as chatbots, will provide new impetus in research on federated search, especially in the context of such RAG pipelines. We show an example of such a RAG pipeline in Figure~\ref{fig:architecture}. Here, a user utterance that has been identified as a request for information is forwarded to information resources so as to trigger a search for relevant information. Retrieved information is then returned to the agent, which uses a Large Language Model (LLM) to aggregate and synthesise the search results, and generate a response for the user.
Example of RAG pipelines that implement this are langchain~\cite{Chase_LangChain_2022}, llamaindex~\cite{Liu_LlamaIndex_2022}, and DSPy~\cite{khattab2022demonstrate,khattab2023dspy}, among others. 

Key tasks in federated search are (a) resource selection, which refers to the selection of a subset of available resources to which forward the search request, and (b) result merging, which refers to the aggregation of the search result lists retrieved from each resource into a unique list of results to be passed onto the response generation process. If multiple resources are to be searched, some of the current RAG pipelines typically employ naive approaches such as the round-robin merging of the top k results from all resources. The absence of resource selection strategies have both quality and cost/latency implications. Obtaining information from irrelevant resources might increase the chances for LLMs to hallucinate~\cite{cuconasu2024power}. But interacting with all resources available (i.e. performing an individual search on each resource), despite many not being relevant, has many other implications. This in fact may attract higher computational and network load, and possible payment of API costs. In addition, there may be higher latency for the merging process (depending on the complexity of the result merging method) and information aggregation/answer generation (assuming that the LLM responsible for this needs to perform inferences for each of the retrieved results). Similarly, the absence of effective result merging approaches also might have implications on the quality of the answers being generated. 

In this paper, we introduce FeB4RAG, a new federated search dataset tailored for integration within the RAG framework. Distinct from existing datasets, FeB4RAG addresses specific limitations previously encountered in federated search collections, which we elaborate upon in Section~\ref{sec:current_limitations}. A key innovation in our dataset is the utilization of advanced LLMs for relevance labelling. We demonstrate that these LLM-generated labels exhibit a high degree of agreement compared to human annotations, thereby ensuring the reliability and applicability of our dataset (Section~\ref{sec:stat}). Furthermore, we detail the steps for constructing the FeB4RAG collection (Section~\ref{sec:method}) and show how to use our collection to evaluate federated search methods and directions for further expansion of the collection (Section~\ref{sec:usage}). 
Lastly, we show the merits of effective federated search in RAG in Section~\ref{sec:demo}. There, we use our FeB4RAG resource to compare a naive federated search strategy and a highly performing one in the context of a RAG pipeline tasked to generate answers to users requested based on information retrieved across multiple resources. The results, displayed in Figure~\ref{fig:winning_cases}, showcase that current naive practices to federated search used in RAG (e.g., round-robin selection) are far from optimal, and thus motivate researching effective methods for federated search in RAG. Our new proposed collection provides a mean to effectively evaluate such methods, addressing the limitations of previous federated search collections that make them not suitable for the considered RAG context.

%% file: sections/methods.tex
  \begin{figure}[t]
	\centering
	\includegraphics[width=\columnwidth]{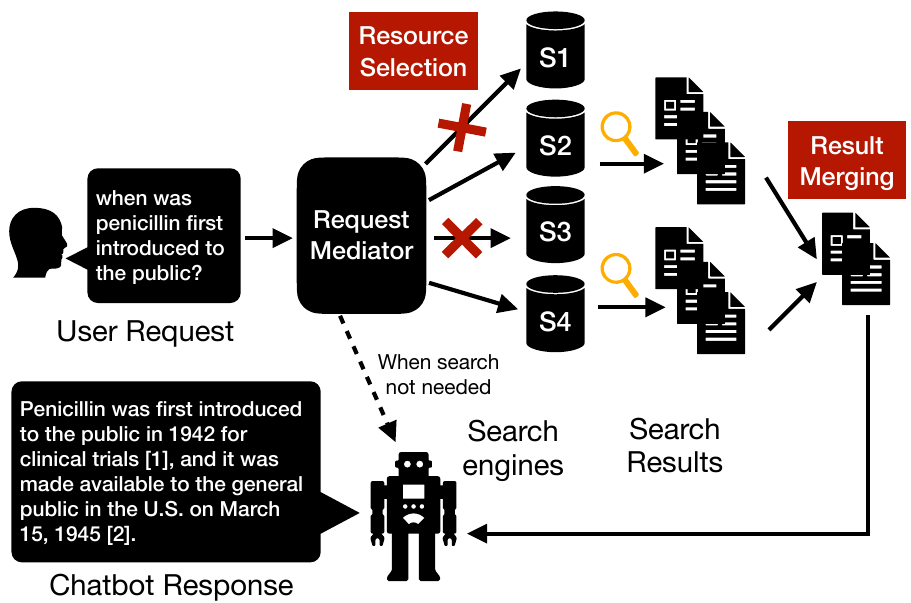}

	\caption{Architecture of Federated Search within RAG.}
	\label{fig:architecture}

\end{figure}

\section{Limitation of Available Federated Search Collections}
\label{sec:current_limitations}
Existing federated search systems can be broadly classified into two primary categories: uncooperative and cooperative~\cite{garba2023federated}. \textit{Uncooperative} federated search systems entail the use of individual search engines that operate independently of the federated search process. These engines and their underlying collections, including statistical data, remain opaque to the federated system. So, for example, collection statistics, such as the frequency of a term in one of the specific resources, are unknown to the federated system. Conversely, \textit{cooperative} federated search systems involve engines or databases that, despite not being originally designed for federated search, offer some level of support for the process, e.g., provide term frequency information. To address these environments, current federated search test collections are divided into Uncooperative Collections and Cooperative Collections. However, existing datasets in both categories exhibit notable limitations that hinder their efficacy as evaluation tools for modern federated search systems, particularly when integrated into RAG frameworks.

\textbf{Uncooperative Collections:} Uncooperative Collections, as exemplified by TREC FedWeb 2013 and 2014~\cite{demeester2014overview,demeester2015fedweb}, predominantly utilize real search engines, obtaining results through API calls. This approach presents two major drawbacks.
Firstly, the search algorithms employed by these search engines are often proprietary and undisclosed. This lack of transparency poses significant challenges in understanding the specific search behaviour of each engine and the rationale behind the retrieval of certain results. 
Secondly, these collections face the risk of obsolescence due to their reliance on active search engine services, which are subject to discontinuation or updates. A notable example is the transition from TREC 2013 to 2014, where eight search engines were removed due to service unavailability within a single year. 
Such instability not only undermines the consistency of these collections but also poses substantial barriers to their future adoption, expansion and adaptability.
This includes challenges when adopting new queries or integrating emerging resource selection technologies, given the reliance of the search engine access to API calls.

\textbf{Cooperative Collections:} Standard web collections, such as ClueWeb09~\cite{callan2009clueweb09} or TREC Gov2~\cite{clarke2004overview}, have been adapted to evaluate federated search methods, for example see \citet{ergashev2023learning}.
 These collections however exhibit other limitations. While they are constructed from a single dataset with sub-collections acting as `search engines', it is important to note that they were not originally designed for federated search applications. Instead, they have been adapted to evaluate federated search methodologies. This adaptation process, however, while valid to evaluate shard selection approaches, falls short in accurately representing the diversity inherent in the real-world federation of search engines. These collections lack the specialization and unique characteristics of varied search platforms, thereby significantly diminishing their utility in mimicking realistic federated scenarios. This directly impacts the efficacy of the evaluation instrument in assessing how federated search systems deal with the complex nuances of real-world information retrieval.

\textbf{Adaptation to Modern Requirements:} Both types of available collections are primarily designed for ad-hoc `Web Search' scenarios. 
This focus diverges from the requirements of modern RAG systems like those implemented by conversational agent, where user requests are contextually richer and more specific. This is for example demonstrated by current federated search collections containing only short keyword queries; while queries from RAG pipelines are likely being more akin to natural language requests.

Additionally, the technology underpinning the search engines in previous collections, rooted in the early 2010s, largely involves ``simpler'' keyword matching or early learning-to-rank methods. In contrast, contemporary search methods, like those emerging from neural information retrieval research~\cite{mitra2018introduction,guo2020deep,trabelsi2021neural,fan2022pre,tonellotto2022lecture,zhu2023large,wang2021bert,li2022interpolate}, 
offer a more sophisticated approach, which our collection employs. 

These limitations highlight the need for developing new federated search collections that align with current technological advances and nuanced user requirements, thereby advancing the field and ensuring the relevance of research to real-world applications.

\section{Dataset Creation}
\label{sec:method}

This section outlines the creation of the FeB4RAG dataset, focusing on search engine selection, user request creation and relevance labelling. Each step is crucial for ensuring the dataset's robustness and applicability in federated search within a RAG framework. Table~\ref{table:datasets} lists the 16 datasets we include in the FeB4RAG collection, along with the corresponding retrieval models that simulate the search engine. Datasets are grouped by retrieval task.

\begin{table}[t]
	\centering

	\caption{Overview of FeB4RAG with respect to each BEIR dataset. \#Req refers to the number of request devised from each BEIR dataset; LLM Req? to whether an LLM was needed to devise requests; Size to the number of documents in the dataset. The abbreviation Ret. stands for Retrieve.}
	\label{table:datasets}
	\input{dataset_table.tex}

\end{table}

\subsection{Search Engine Selection}
\label{sec:search_engine_selection}
In our collection, we utilize existing datasets, each paired with a corresponding state-of-the-art retrieval model, to function as the search engines. 
Our selection process is guided by the BEIR benchmark, widely recognised  for IR  evaluation~\cite{thakur2021beir}. 
This benchmark's datasets form the foundational basis for our search engine selection.
The next step involves identifying effective retrieval models for each of these datasets to be used to simulate the individual search engines. To guide this process, we use the MTEB leaderboard~\cite{muennighoff2022mteb}, 
that tracks performance of state-of-the-art embedding models. Most BEIR collections are included in the leaderboard\footnote{Except those that are not public, such as Signal-1M, TREC-NEWS and Robust04.}.

Of the 15 BIER datasets that are also present in the MTEB leaderboard, we exclude Quora and CQADupStack due to their focus on duplicate question retrieval, which does not align with our focus on tasks likely to emerge from users information requests in RAG systems such as conversational agents. For the remaining 13 datasets, we only considered those top performing models that are based on dense retrievers, because dense retrievers are generally characterised by low latency and computational requirements, and so we could easily scale and run. Note that we excluded models based on APIs, models for which no source-code was provided, and models with more than 6 billion parameter -- again due to computational and cost constraints. 
Each dataset, in conjunction with its designated retrieval system, then serves as a simulated search engine in our federated search collection.

In addition, we include three other datasets from the BEIR collection: TREC-News, Robust04, and Signal1M. To select which dense retriever to use as the search engine on these datasets, we evaluate the top 30 dense retrievers from the MTEB leaderboard, 
 selecting the best-performing model for each dataset\footnote{Evaluated by nDCG@10 on the datasets' test portions.}. 

All model choices were based on the effectiveness reported on MTEB as of 31/12/2023; models released after this date are not included in our selection process. As a result, our final dataset comprises 16 BEIR datasets, each served through a specific retrieval model. While this number is considerably less than the number of search engines used within the TREC FedWeb collections, we see this fitting many real application of RAG pipelines in the context of conversational agents (e.g., within enterprise settings).
Table~\ref{table:datasets} enumerates the retrieval models used for each of the BEIR benchmark's 16 datasets, spanning 8 tasks and 8 verticals\footnote{Vertical refers to the domain of the resource~\cite{demeester2014overview}.}. In total, 9 distinct retrieval models are employed.

\subsection{User Requests Creation}
The presence of user requests likely to be used within RAG applications is a pivotal aspect of the FeB4RAG dataset. We employed a structured approach to generate high-quality, diverse information request that fit the datasets we included in the collection. The methodology encompasses the following steps:

\begin{itemize}[leftmargin=*]
	\item[\textbf{(1)}] \textbf{Initial Query Assessment:} Our process commenced with an evaluation of the original test queries from the selected BEIR datasets. We aimed to select roughly 50 queries per dataset, to comply with our computational and assessment budgets. Notably, many queries, especially argumentative ones from ArguAna~\cite{wachsmuth2014a-review} or paper titles from SCIDOCS~\cite{cohan2020specter}, were deemed unsuitable for RAG settings in their original form.
	
	\item[\textbf{(2a)}] \textbf{Manual Request Selection:} For datasets inherently comprising question-based queries with clearly articulated user information needs, we conducted a manual review and selection process. For each dataset, we first ordered queries in descending order of the number of associated relevant documents. Then, we went down the list and selected queries that were deemed to be appropriate user requests likely to be used within the targeted RAG pipelines. 
	 During this selection, we excluded requests that lacked specific information needs, were ambiguous or confusing, too short and overly simplistic\footnote{For our criteria of simplicity, queries consisting of four terms or less were deemed too simplistic and thus omitted}.

	\item[\textbf{(2b)}] \textbf{Objective Identification and Prompt Creation:} For datasets lacking explicit information needs in all queries, we identified the core objective of each dataset. Based on these objectives, we developed specific prompts for a Large Language Model (LLM) to facilitate the formulation of queries into a conversational information request style. Table~\ref{table:datasets} reports for which datasets we resorted using LLMs to create requests. Table~\ref{table:query_prompt} illustrates the prompts utilized for each relevant dataset during this process.
	
	\item[(\textbf{3b})] \textbf{Query Rewriting using GPT-4:} Relying on the devised prompts, we used GPT-4 to transform the original queries contained in the affected datasets to information requests likely to be found in RAG applications. 
	To promote the inclusion of diverse user requests formats, especially for datasets in which all queries have a similar objective, we implemented a looping strategy. This involved generating multiple user requests variations using GPT-4 in a multi-step process, building upon previous LLM inputs. For this user request generation phase, we used the \texttt{gpt-4-0125-preview} model with $temperature=0$ and $seed=1$. We set temperature to zero to allow for replicability.
	
	\item[(\textbf{4b})] \textbf{Manual Request Selection of Rewritten Requests:} The concluding phase involved a meticulous manual selection process of LLM-generated information requests. We prioritized requests that displayed a diverse range of expressions of the user information needs for analogous tasks. This selection was critical in ensuring the richness and diversity of the final user requests.
\end{itemize}

Finally, our methodology resulted in the generation of 790 user requests to be included in FeB4RAG. The distribution of these user requests, derived from each dataset, along with their categorization, is detailed in Table~\ref{table:datasets}. Notably, we encountered limitations in achieving the target of 50 queries for two datasets. Specifically, for Touché-2020~\cite{bondarenko2020overview}, we could only include 49 queries as the test set itself contained only 49 queries. Similarly, in the case of MS MARCO~\cite{nguyen2016ms}, we incorporated 41 queries only because a significant number of queries were excluded during manual request selection.

\begin{table}
	\centering

	\caption{For datasets that only contained queries deemed unsuitable of typical RAG pipelines applications, we created prompts for an LLM to generate queries from one of the queries of the dataset. The table reports the prompt used for each of the affected datasets.}
	\label{table:query_prompt}
	\input{query_prompt_table.tex}

\end{table}

\subsection{Relevance Labelling}
Relevance evaluation in a federated search collections involves two stages: (1) assessing the correct selection of the search engine, and (2) evaluating the relevance of search results retrieved from each search engine.  In constructing our collection, we first determined the relevance of individual search results, which were then aggregated to establish search engine level relevance. This follows common practice in previous TREC FedWeb collections.

Note that in creating the relevance labels for our collection, we could not rely on the original relevance judgements from the BEIR benchmark, due to several reasons:

\begin{enumerate}
	\item Sparse judgments in datasets like ArguAna and MS MARCO, leading to many unjudged top documents.
	\item Slight changes in query contexts due to LLM-based rewriting of test queries into user information requests.
	\item Absence of annotations for results from search engines not aligned with the original user requests.
	\item Varied scales of relevance judgments in the labels from the BEIR datasets.
\end{enumerate}

\subsubsection{Search Result Preparation}
To initiate the relevance judgment process for search results, we ran the user requests onto each identified search engine, as delineated in Section~\ref{sec:search_engine_selection}. We then retrieved the top-$k=1$ results from each search engine for further analysis.

\subsubsection{Labelling Search Results}
Recent work has considered the use of LLMs for creating reliable relevance judgements without human intervention~\cite{faggioli2023perspectives,thomas2023large}. In particular, 
\citeauthor{thomas2023large} have demonstrated that GPT-4 can effectivelly produce relevance judgments that closely align with those of human annotators~\cite{thomas2023large}. We adopt their approach, modifying their most effective prompts to suit our specific needs.

As outlined in Table~\ref{table:label_prompt}, we categorize relevance into four levels: key (score=3), high relevance (score=2), minimal relevance (score=1), and not relevant (score=0), paralleling the TREC FedWeb datasets~\cite{demeester2014overview,demeester2015fedweb}. However, we omitted the `nav' label due to its similarity with `key' in relevance signal in TREC FedWeb. The description of each label we include in the prompt is informed by a recent study that developed prompts for evaluating search results in RAG systems~\cite{gienapp2023evaluating}.

However, employing GPT-4 for relevance judgement incurs substantial costs\footnote{We estimated a cost of approximately USD \$2000 for using the GPT-4 API service to judge the top-10 results from all resources, amounting to 7,901,016 judgements.}. As an alternative to this, we identified two high-performing models from the open LLM leaderboard~\footnote{\url{https://huggingface.co/open-llm-leaderboard}. This leaderboard evaluates the effectiveness of generative Large Language Models across multiple natural language processing tasks.}:

\begin{itemize}[leftmargin=*]
	\item  \texttt{upstage/SOLAR-10.7B-Instruct-v1.0}~\cite{kim2023solar}: a 10.7 billion parameters LLM, which relies on depth up-scaling (DUS): Mistral 7B weights were integrated into the upscaled layers and continued pre-training was adopted for the entire model. It also uses state-of-the-art instruction fine-tuning methods including supervised fine-tuning and direct preference optimization (DPO)~\cite{rafailov2023direct}.

	\item \texttt{RubielLabarta/LogoS-7Bx2-MoE-13B-v0.1} (\texttt{lgs-13b})\footnote{\url{https://huggingface.co/RubielLabarta/LogoS-7Bx2-MoE-13B-v0.1}}: a 13 billion parameters LLM, which relies on a mixture of expert using the spherical linear interpolation merge method. 
\end{itemize}

Our use of these two open LLMs to generate relevance labels resulted in two distinct labels for each search result. We then subsequently aggregated these labels to produce a final label. The final judgment label for each result was determined by averaging the integer number (i.e., score) associated with the label produced by each model (e.g. not relevant is 0, key is 3), and then rounding down to the nearest integer. 
This method mirrors human annotation scenarios where multiple annotators are used for a single document to obtain a final label through aggregation and adjudication. This process typically enhances the reliability and objectivity of the relevance assessments.

%


\subsubsection{Labelling Search Engines}
Search engine level labels are derived from aggregated search result labels using graded precision scores, following TREC FedWeb practice~\cite{demeester2014overview,demeester2015fedweb}. The graded precision score for a user request $r$ with respect to a search engine $s$ is calculated as:

\begin{equation}
	\text{Graded Precision(r, s)} = \left( \frac{\sum_{i=1}^{10} w_{i}(r, s)}{10} \right) \times 100
\end{equation}
where the weight for each search result level relevance is: \( w_{\text{not\_relevant}} = 0 \), \( w_{\text{minimal\_relevance}} = 0.25 \), \( w_{\text{high\_relevance}} = 0.5 \), and \( w_{\text{key}} = 1 \).
After this, the computed relevance score for a search engine ranges from 0 to 100, with 100 indicating the highest relevance to the information request.




\begin{table}
	\centering
	\caption{Prompt used for obtaining relevance judgements from the employed LLMs.}
	\label{table:label_prompt}
		\begin{tabular}{p{2pt}|M{410pt}}
			\toprule
			& \textnormal{Prompt} \\
			\hline
			\rotatebox[origin=c]{90}{System Prompt} & Given a user request and a search result, you must provide a score on an integer scale of 0 to 3 with the following meanings: \linebreak 3 = key, this search result contains relevant, diverse, informative and correct answers to the user request; the user request can be fulfilled by relying only on this search result. \linebreak  2 = high relevance, this search result contains relevant, informative and correct answers to the user request; however, it does not span diverse perspectives, and including another perspective can help with a better answer to the user request. \linebreak 1 = minimal relevance, this search result contains relevant answers to the user request. However, it is impossible to answer the user request based solely on the search result.  \linebreak0 = not relevant, this search result does not contain any relevant answer to the user request. \\ \midrule
			
			\rotatebox[origin=c]{90}{{\centering User Prompt}} & Assume that you are collecting all the relevant search results to write a final answer for the user request. \linebreak User Request: \linebreak A user typed the following request. \linebreak \{request\} \linebreak Result:\linebreak Consider the following search result: \linebreak —BEGIN Search Result CONTENT—\linebreak \{result\}\linebreak —END Search Result CONTENT—\linebreak Instructions:\linebreak Split this problem into steps:\linebreak Consider the underlying intent of the user request. \linebreak Measure how well the content matches a likely intent of the request (M)\linebreak Measure how trustworthy the search result is (T).\linebreak Consider the aspects above and the relative importance of each, and decide on a final score (O).\linebreak Produce a JSON of scores without providing any reasoning. Example:\{"M": 2, "T": 1,"O": 1\}\linebreak Results:\\
			\bottomrule
		\end{tabular}
\end{table}

%% file: dataset_table.tex
\begin{tabular}{l|l|l|l|lll}
	\toprule
	\textbf{Task} & \textbf{Vertical} &  \textbf{Dataset} & \textbf{Selected Model} & \textbf{\# Req} & \textbf{LLM Req?} & \textbf{Size} \\
	\midrule
	Ret.Passage& General & MS MARCO & e5-large~\cite{wang2022text}  & 41 & \xmark &  8.8M \\ \midrule
	
	\multirow{3}{*}{Ret.Publication} & Biomedical & TREC-COVID & SGPT-5.8B~\cite{muennighoff2022sgpt}  & 50 & \xmark & 171K \\
	& Biomedical & NFCorpus & UAE-Large-V1~\cite{li2023angle}  & 50 & \xmark & 3.6K \\
	& Scientific & SCIDOCS & all-mpnet-base-v2~\footnote{\url{https://huggingface.co/sentence-transformers/all-mpnet-base-v2}}  & 50 & \cmark & 26K  \\ \midrule
	\multirow{3}{*}{Q\&A}& Wiki & NQ & multilingual-e5-large~\cite{wang2022text}  & 50 &  \xmark & 2.6M  \\
	& Wiki & HotpotQA & ember-v1~\footnote{\url{https://huggingface.co/llmrails/ember-v1}} & 50 & \xmark & 5.2M \\
	& Finance & FiQA-2018 & all-mpnet-base-v2~\footnote{\url{https://huggingface.co/sentence-transformers/all-mpnet-base-v2}}  & 50 & \xmark &  58K\\
	\midrule
	Ret.Tweet & Tweet & Signal-1M & gte-base~\cite{li2023towards}  & 50 & \cmark & 2.8M  \\
	\midrule
	\multirow{2}{*}{Ret.News}& News & TREC-NEWS & gte-large~\cite{li2023towards}  & 50 & \cmark &   595K \\
	 & News & Robust04 & instructor-xl~\cite{su2022one}  & 50 & \xmark  & 528K \\
	\midrule 
	\multirow{2}{*}{Ret.Argument}& General & ArguAna & UAE-Large-V1~\cite{li2023angle}    & 50 & \cmark  & 8.6K \\
	& Debate & Touché-2020 & e5-base~\cite{wang2022text}  & 49 & \xmark  &  383K \\ \midrule
	Ret.Entity & General & DBPedia & UAE-Large-V1~\cite{li2023angle}    & 50 & \cmark & 4.6M  \\ \midrule
	\multirow{3}{*}{Fact Checking} & Wiki & FEVER & UAE-Large-V1~\cite{li2023angle}    & 50 & \cmark & 5.4M  \\
	&Wiki & ClimateFEVER  & ember-v1  & 50 & \cmark &  5.4M  \\
	& Scientific & SciFact & gte-base~\cite{li2023towards} & 50 & \cmark & 5K  \\ \midrule
	
	Overall && FeB4RAG&& 790& & 36.9M \\ \bottomrule
\end{tabular}
\vspace{-0.3cm}

%% file: query_prompt_table.tex
\begin{tabular}{l|l|M{300pt}}
	\toprule
	& Dataset & \textnormal{Prompt} \\
	\midrule
	\multicolumn{2}{l|}{System Prompt}  & You are an helpful assistant helping to reformulate the provided text, that describes a need of a user, into a conversational question that expresses the user need. The generated text will be used for a user to ask a chatbot for a direct response.
	Therefore, it should not include information about the retrieval step.\\ \midrule
	\multicolumn{2}{l|}{User Prompt start} &  Consider the following text. \\ \midrule
	
	 \multirow{20}{*}{User Prompt} & SCIDOCS  & The text is the title of a research article. I want you to formulate a question that asks to find related articles to the one provided in the text. \\  \cmidrule{2-3}
	 &Signal-1M  & The text is the title of a news article. I want you to formulate a question that asks to find relevant Tweet messages about the provided news article. In your question, do not mention that the text is the title of the news article.\\  \cmidrule{2-3}		
	 &TREC-NEWS&The text is a topic, The text is a topic, I want you to formulate a question that asks to find relevant news based on the topic. \\ \cmidrule{2-3}
	 &ArguAna& The text provides an argument with claims. I want you to formulate a question that asks to find counter-arguments to the main claim in the text. In your question, specify clearly what that claim is, but do not refer explicitly to the text. \\ \cmidrule{2-3}
	 &DBPedia& The text provides an entity. I want you to formulate a question that asks to find relevant information about the entity. In your question, do not mention that the text is an entity. \\ \cmidrule{2-3}
	 & (Climate-)FEVER& The text is a claim. I want you to formulate a question that asks to find evidence that supports or refutes the claim made in the text. In your question, do not specify that you want to find evidence. \\ \cmidrule{2-3}
	 &SciFact &  The text is a scientific claim. I want you to formulate a question that asks to find evidence that supports or refutes the claim made in the text. In your question, do not specify that you want to find evidence.\\ \midrule
	 \multicolumn{2}{l|}{User Prompt end} & Please only include the formulated question in your response. \linebreak
	 Text:\linebreak \{query\} \\ \midrule
	 \multicolumn{2}{l|}{User Prompt Loop} & Please formulate again so that it is different from the previous response. \\ 
	 
	\bottomrule 
\end{tabular}

%% file: sections/experimental_setup.tex
\section{Analysis of Relevance Labelling}
\label{sec:stat}
This section delves into a comprehensive analysis of relevance labelling within the FeB4RAG collection. This is important because we break with common practice in IR, and use LLMs rather than human annotators to obtain relevance assessments. 
Our focus encompasses several key areas: the overall labelling statistics, agreements between labels generated by LLMs and human annotations, cross-LLM labelling agreements, and the importance of each search engine vertical as inferred from the engine-level labels.

\begin{table}[t!]
	
	\caption{Statistics with respect to annotations from Feb4RAG with respect to search result level and engine-level.}
	\begin{center}
			\begin{tabular}{l|llll|ll}
				\toprule
				Search Engine& \multicolumn{4}{c|}{Result Level (\%)}  & \multicolumn{2}{c}{Engine Level}   \\  \cmidrule{2-7}
				Collection & Key & HR & MR & NR & Avg & Max \\ \midrule
				MS MARCO & 1.41  & 10.91  & 52.15  & 35.53  & 20.14 & 90 \\ \midrule
				TREC-COVID & 0.16  & 4.84  & 22.82  & 72.18  & 8.48 & 56 \\ \midrule
				NFCorpus & 0.01  & 2.14  & 13.75  & 84.10  & 4.63 & 45 \\ \midrule
				SCIDOCS & 0.05  & 2.04  & 25.84  & 72.08  & 7.72 & 53 \\ \midrule
				NQ & 0.28  & 10.34  & 43.38  & 46.00  & 16.52 & 60 \\ \midrule
				HotpotQA & 0.41  & 6.44  & 41.59  & 51.56  & 14.27 & 58 \\ \midrule
				FIQA-2018 & 0.06  & 1.33  & 19.63  & 78.97  & 5.82 & 43 \\ \midrule
				Signal-1M & 0.00  & 1.01  & 29.04  & 69.95  & 7.97 & 48 \\ \midrule
				TREC-NEWS & 0.22  & 11.23  & 53.63  & 34.92  & 19.48 & 65 \\ \midrule
				Robust04 & 0.04  & 4.08  & 43.86  & 52.03  & 13.24 & 40 \\ \midrule
				ArguAna & 0.01  & 1.29  & 18.53  & 80.16  & 5.42 & 40 \\ \midrule
				Touché-2020 & 0.08  & 2.16  & 28.18  & 69.58  & 8.39 & 48 \\ \midrule
				DBPedia  & 0.41  & 5.24  & 42.68  & 51.67  & 13.92 & 53 \\ \midrule
				FEVER & 0.99  & 11.77  & 47.84  & 39.41  & 19.08 & 80 \\ \midrule
				Climate-FEVER & 0.97  & 11.90  & 47.76  & 39.37  & 19.11 & 80 \\ \midrule
				SciFact & 0.05  & 2.39  & 15.73  & 81.82  & 5.31 & 48 \\ \midrule \midrule
				Overall & 0.32  & 5.57  & 34.15  & 59.96  & 11.84 & 90 \\ \bottomrule
				
			\end{tabular}
	\end{center}

	\label{table:label_statistics}
\end{table}
\subsection{Labelling Statistics}

Table~\ref{table:label_statistics} summarises the primary statistics with respect to our relevance labelling. 
We note a predominance of search results classified as either minimally relevant or not relevant. Key and high relevance labels are relatively rare, averaging at 0.32\% and 5.57\%, respectively. This trend suggests the LLMs may have followed a stringent labelling criterion, ensuring that only the most pertinent results are categorized as highly relevant.

Upon comparing relevance labels at the search result level across different search engines, we observed distinct patterns based on the corpus size. Smaller scale corpora, such as NFCorpus (3.6k), SciFact (5k), and ArguAna (8.67k), exhibit a higher prevalence of non-relevant assessments, with 84.10\%, 81.82\%, and 80.16\% of the search results being judged as not relevant, respectively. This trend highlights the challenge in extracting relevant information from smaller datasets, and the likely misalignment of some of the resources with a portion of the information requests in the collection. This characteristic is likely found also in real applications where federated search might be used.

In terms of engine-level relevance scores, a similar trend emerges. Larger collections, like MS MARCO, demonstrate a higher chance of being relevant collections to most user information requests, compared to smaller collections. We highlight that this may not only be due to the size of the collection but also to the actual nature of the collection itself (i.e. a broad, general collection, compared to focused collections such as FEVER).

\begin{figure*}[t]
	\centering
	\subfigure[solar-11b]{
		\includegraphics[width=0.3\textwidth]{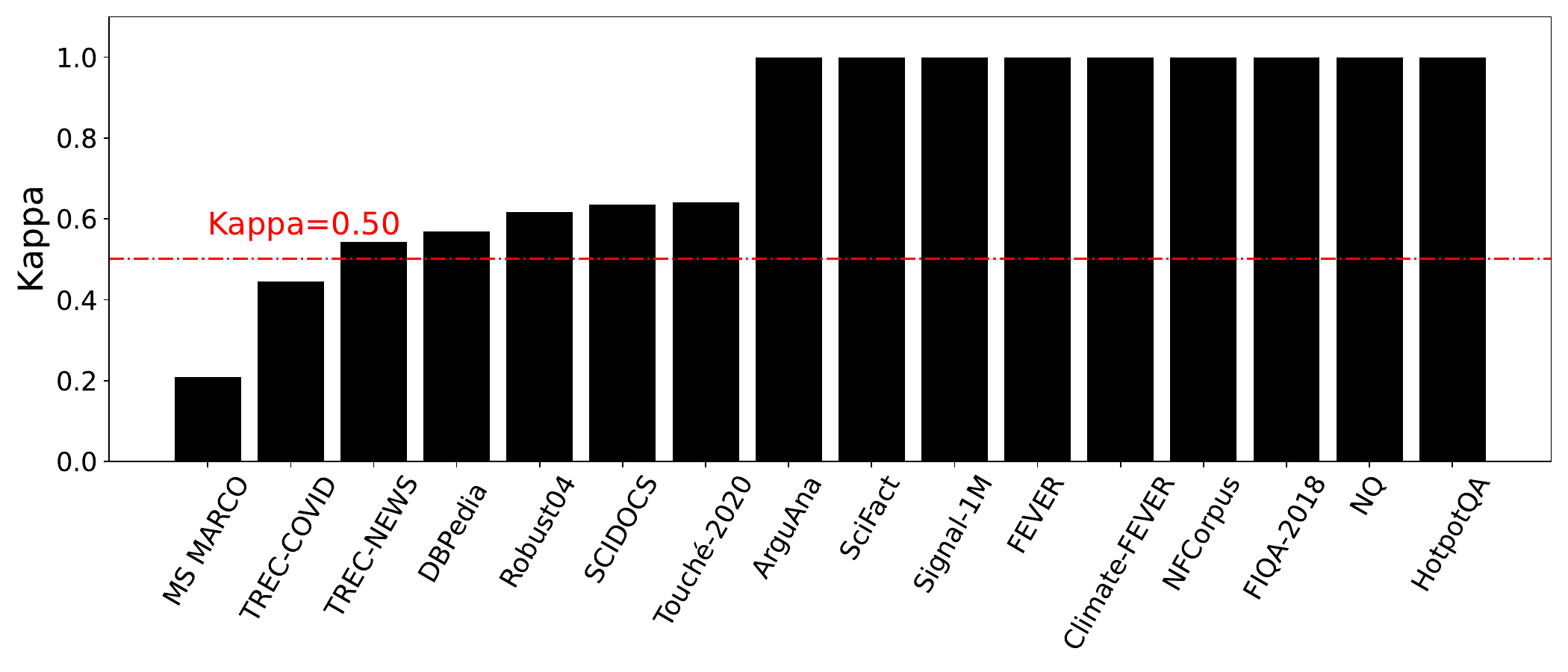}
		\label{fig:solar-11b}
	}
	\subfigure[lgs-13b]{
		\includegraphics[width=0.3\textwidth]{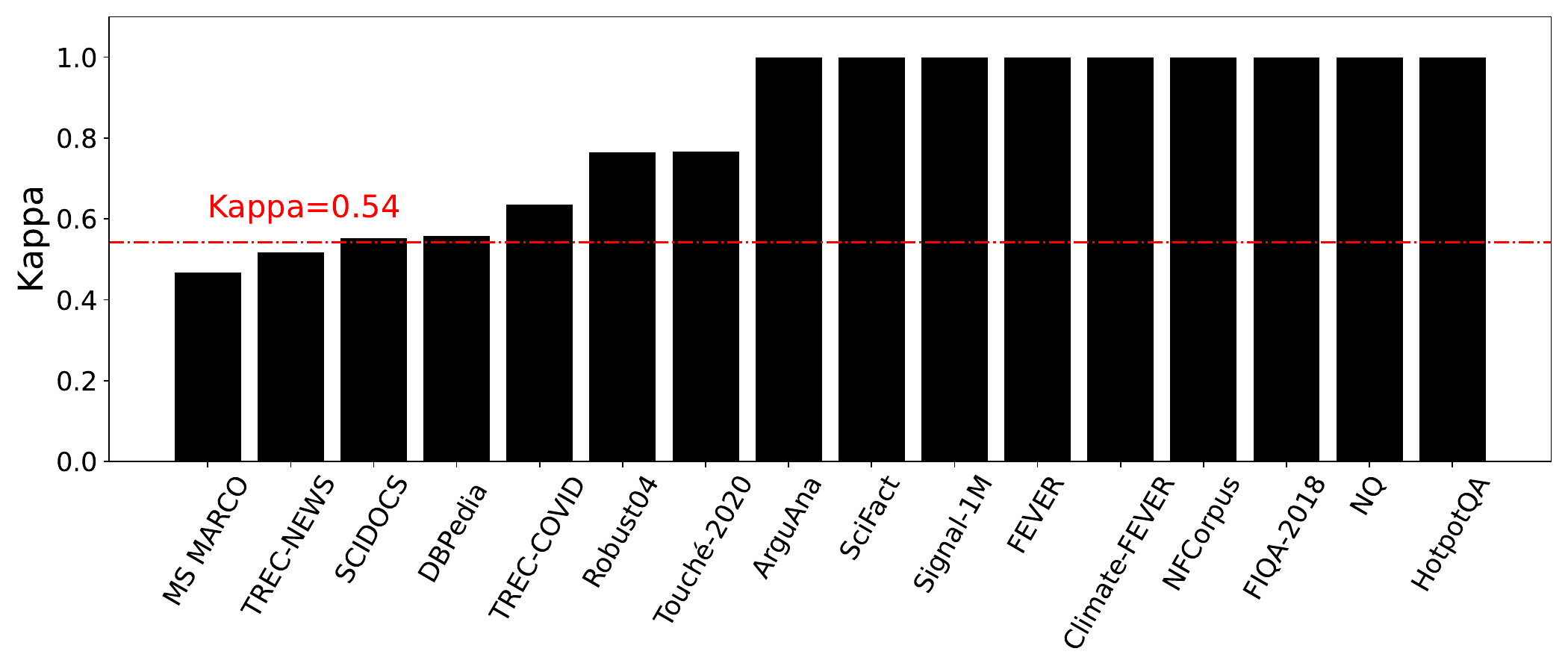}

		\label{fig:lgs-13b}
	}
	\subfigure[Fused (solar-11b + lgs-13b)]{
		\includegraphics[width=0.3\textwidth]{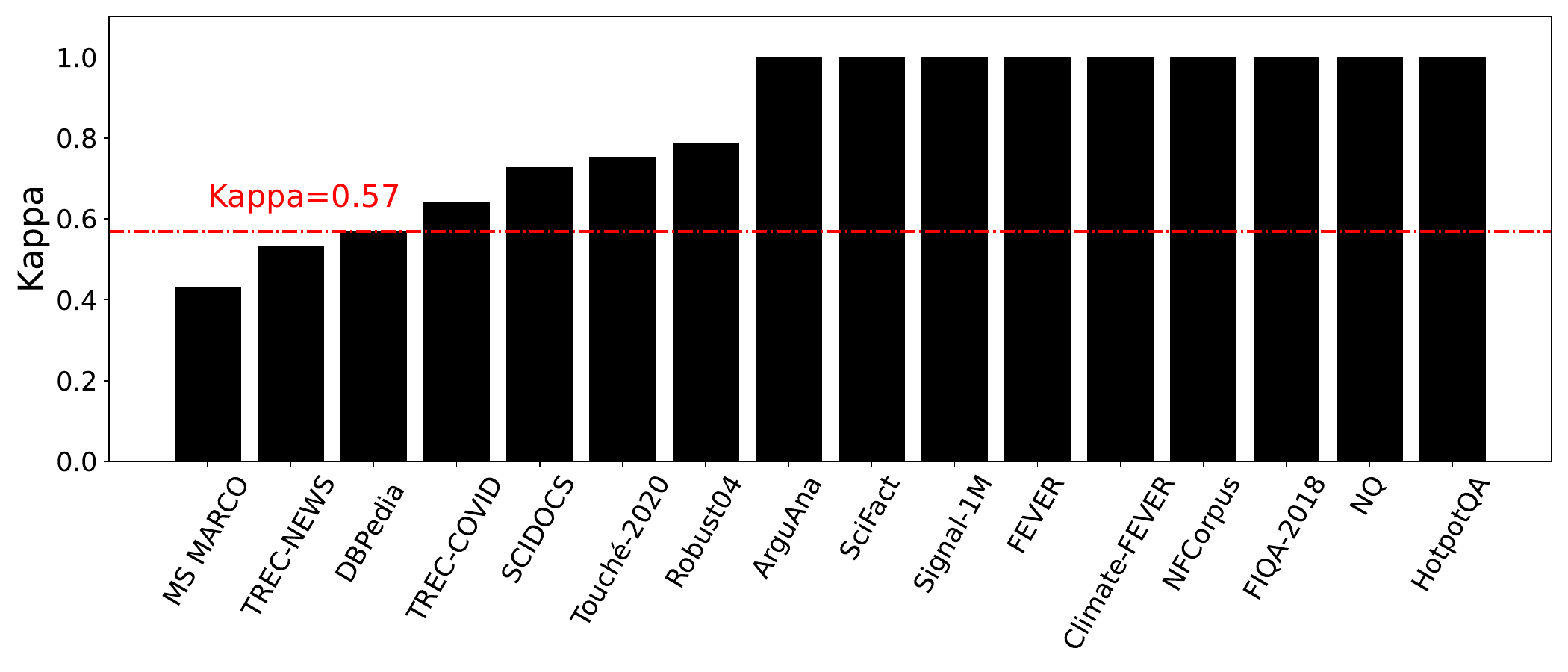}
		\label{fig:fused}
	}

	\caption{Cohen's Kappa between labels generated by the LLM and labels provided by humans (from the original datasets); red line indicates overall Kappa for all annotations. }
	\label{fig:preliminary_label}
\end{figure*}

\subsection{LLM-based Labels vs. Human Annotations}
Since the relevance labels in our experiments are derived from two LLMs (solar-11b and lgs-13b), an essential aspect of our collection's integrity is verifying the agreement between LLM-derived relevance labels and human annotations. For this purpose, we conducted an experiment using the two LLMs to validate their reliability in relevance judgment. 

To ensure a comprehensive evaluation of the agreement between LLM-derived labels and human annotations, our methodology involved a systematic selection process. For each of the 790 user requests in our FeB4RAG collection, which were derived from the queries in the BEIR dataset, we randomly selected one relevant and one irrelevant search result. These selections were based on the human annotations provided in their corresponding BEIR datasets. It is important to note that for those collections within BEIR where only relevant judgments are provided, our selection was constrained to these relevant judgments only.

Following this, the selected search results were labelled using the two LLMs. To facilitate a meaningful comparison with the original BEIR datasets — which employed heterogeneous labelling scales, including binary, 3-level, and 5-level scales — we binarized the LLM labels into two classes: relevant (graded label > 0) and not relevant (graded label = 0). This binarization was crucial to align our 4-level labelling scale to the diverse scales of the original datasets. 

We then calculated Cohen's Kappa for each collection to determine the level of agreement between the LLM judgments and human annotations. The results, illustrated in Figure~\ref{fig:preliminary_label}, show a substantial level of agreement between the two. All datasets achieved a Cohen's Kappa above 0.4, which indicates moderate agreement. Notably, the aggregated labels from both LLMs often resulted in the highest average Kappa, 0.57, compared to 0.54 when using lgs-13b and 0.5 when using solar-11b. This underlines the efficacy of our LLM-based labelling approach based on the aggregation of the labels produced by the two separate LLMs.

These results enhance our confidence in the reliability of LLM-based relevance judgments, demonstrating their considerable similarity to human-annotated judgments. This experiment substantiates our methodology of employing LLM-judged results for relevance assessment in the FeB4RAG collection.

\subsection{Agreements between LLMs}

Next, we assess the agreement level with respect to the relevance labels between the two LLMs, solar-11b and lgs-13b. For this, we used the same scale of relevance for both LLMs, thereby obviating the need for binarization in the computation of Cohen's Kappa, differently from the previous section. The agreement was computed across all search results from all search engines, based on the 790 user requests in our collection. This amounted to a substantial total of 7,901,610 individual comparisons.

\begin{figure}[t]
	\centering
	\includegraphics[width=\textwidth]{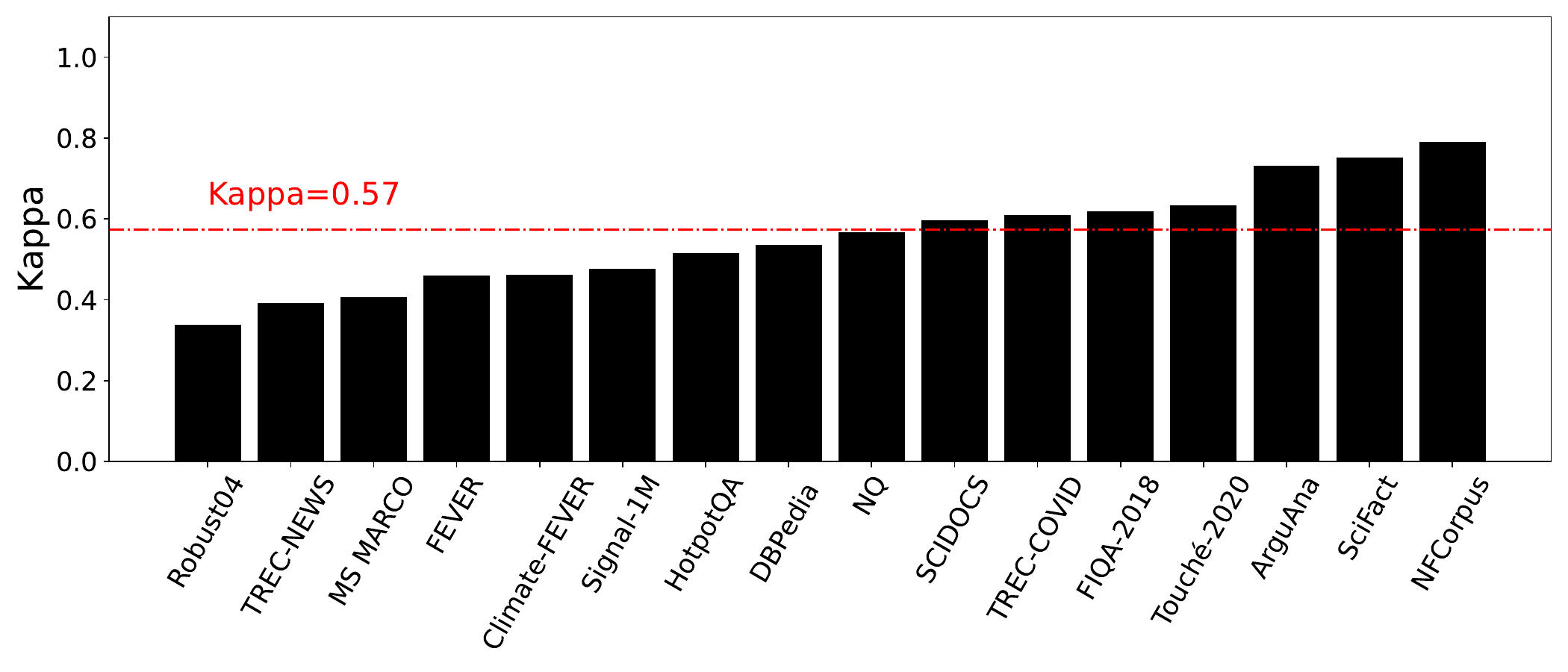}

	\caption{Cohen's Kappa between two LLM annotators: solar-11b and lgs-13b; red line indicates overall Kappa for all annotations.}
	\label{fig:label_two_judge}
\end{figure}

Figure~\ref{fig:label_two_judge} reports the Cohen's Kappa results, indicating the level of agreement between the two LLMs. The overall Kappa stood at 0.57, signifying a moderate level of agreement. This degree of agreement is noteworthy, especially when considered in the context of human annotator agreements. For instance, in the Trec FedWeb 2013 collection, a Cohen's Kappa of 0.6 was reported between two human annotators~\cite{demeester2013overview, wang2024resllm}. The fact that the LLMs we used agreed with each other almost as much as human annotators do in similar contexts reinforces the reliability of our LLM-based annotations; it also highlights the potential viability of using LLMs in other relevance assessment tasks~\cite{faggioli2023perspectives}.

\subsection{Importance of each Resource}
\begin{figure}[t]
	\centering
	\includegraphics[width=\textwidth]{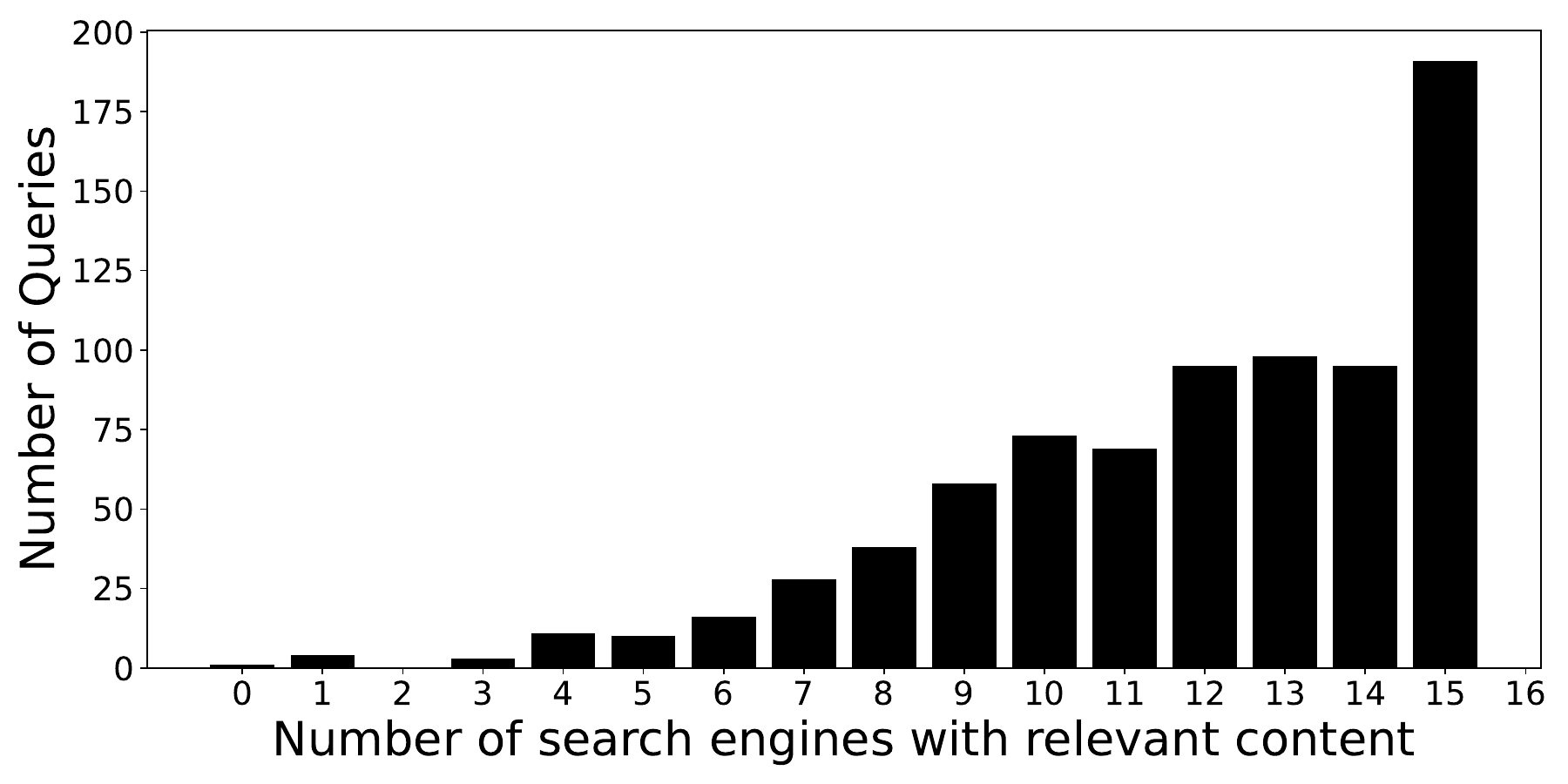}

	\caption{We report the number of queries for which $n$ search engines contain relevant information; we vary $n$ from not to 16.}
	\label{fig:q_res_distribution}
\end{figure}

In federated search, queries are routed to a subset of relevant search engines. But, how many search engines (i.e. BEIR datasets) are relevant for each given query in ReB4RAG? We analyse this aspect in Figure~\ref{fig:q_res_distribution}. We find that on average, for each query there are 11.9 search engines that contain relevant content, and that the collection only contains a handful of queries for which there is just one resource that contains relevant information. 

\subsection{Importance of Resource Vertical}
To understand how each vertical contributes across user requests in FeB4RAG, we plot the graded precision of the best resource vertical for each query. This entails selecting the highest graded precision resource for each user request and then categorizing this resource according to its specific vertical, mapped out in Table~\ref{table:datasets}.


\begin{figure}[t]
	\centering
	\includegraphics[width=\columnwidth]{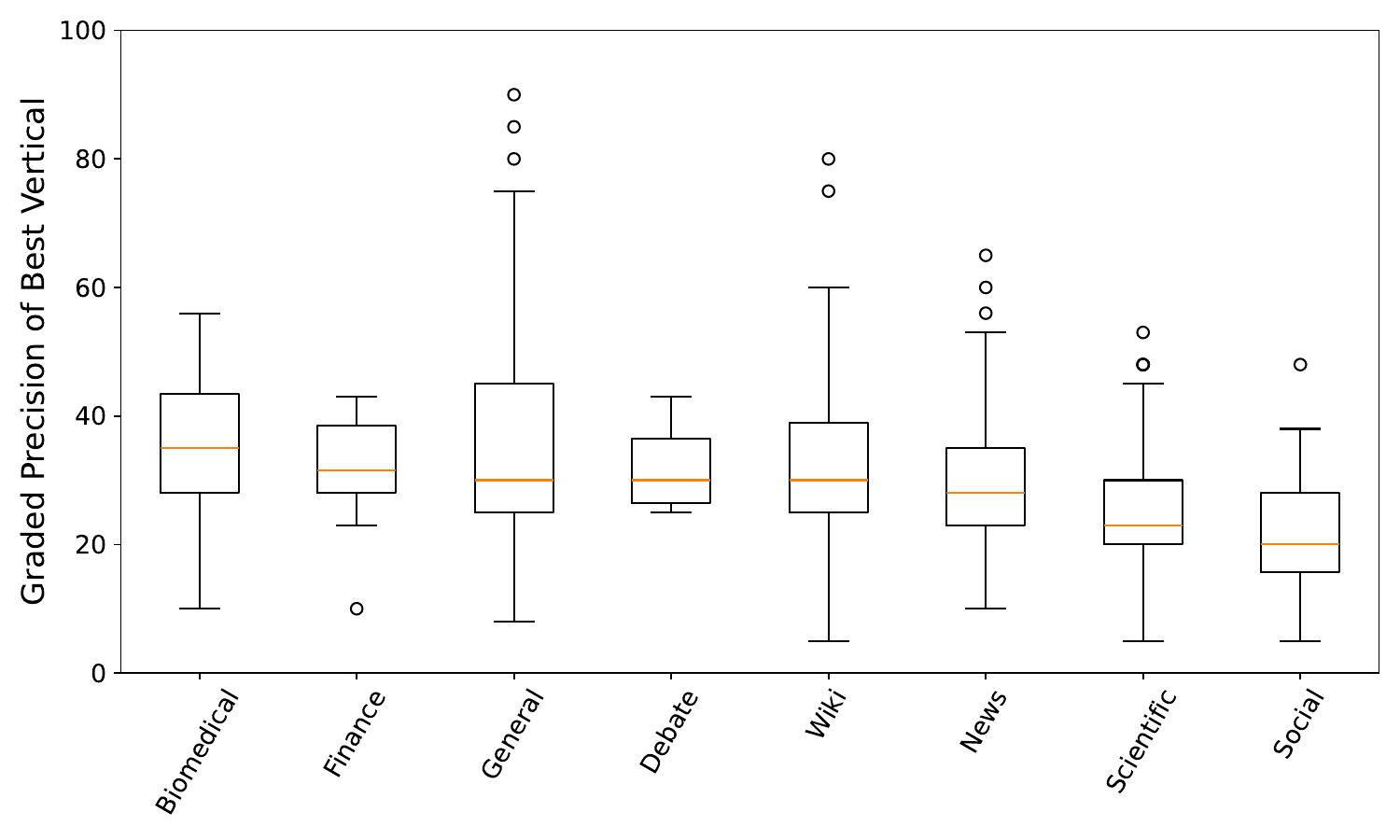}
	
	\caption{Highest graded precision among all resources within a vertical, over 790 user requests. }
	\label{fig:label_resource_importance}
\end{figure}

Figure~\ref{fig:label_resource_importance} displays the variance in importance for each vertical in our collection. A striking observation is the contrast in vertical distribution trends between FeB4RAG and the TREC FedWeb 2014 collection. In FeB4RAG, we observe a more balanced distribution of resource verticals compared to FedWeb. This equitable distribution is significant as it fosters a fairer testing ground for federated search methodologies, at least with respect to verticals. For instance, in previous collections, a resource selection method that prioritized resources based solely on their vertical could inadvertently bias the results but go undetected in the evaluation. Our collection, with its more uniform vertical representation, mitigates this risk, thereby ensuring a more equitable evaluation of federated search methods.

%% file: sections/collection_use.tex
\section{Use of the collection}
\label{sec:usage}
Next we delineate the practical applications of FeB4RAG, highlighting its utility to evaluate federated search tasks like resource selection and result merging. Additionally, we explore potential avenues for expanding the collection to encompass new tasks that emerge from situating the collection within the RAG pipeline.

\subsection{Resource Selection}

Resource selection is a pivotal task in federated search. It involves choosing the most appropriate resources in response to a query. Our FeB4RAG collection facilitates this process by allowing for the ranking or classification of the 16 included search engines. 

The methodology for resource selection may vary depending on the specific approach being developed. In an uncooperative setting, the selector typically does not have access to the underlying corpus of each search engine, necessitating reliance on other factors like the vertical and task associated with each search engine~\cite{garba2023federated}. 
 Approaches in this setting include sample-based methods, which obtain a document sample from the search engine either through query logs, or by issuing single-term queries. Examples of unsupervised methods in this category are ReDDE~\cite{si2003relevant}, CRCS~\cite{shokouhi2007central}, SUSHI~\cite{thomas2009sushi}, and ResLLM~\cite{wang2024resllm}, alongside supervised methods like SVMrank~\cite{dai2017learning} and SLAT-ResLLM~\cite{wang2024resllm}.
 Conversely, in a cooperative setting, selectors can utilize the underlying corpus of each search engine, potentially leading to more informed and effective resource selection~\cite{garba2023federated}. Methods in this setting often involve Lexicon-based approaches that leverage collection statistics based on terms in the collection, as developed by~\citet{callan2002distributed,xu1999cluster}.
Common evaluation metrics employed in TREC FedWeb are applicable to our FeB4RAG collection as well, such as nDCG@k and nP@k~\cite{demeester2013overview, demeester2014overview}.


\subsection{Result Merging}
Result merging is another key task in federated search, involving the integration of results from multiple search engines into a coherent and effective response. Common techniques for result merging include the naive execution of a round-robin merging (this is often the approach deployed as default in open-source RAG pipelines), and more complex result fusion techniques~\cite{tjin2010learning,mourao2013novasearch}. In the context of FeB4RAG, the effectiveness of result merging is contingent upon the effectiveness of the resource selection process.

Common evaluation measures used in TREC FedWeb, such as nDCG@k, can also be used on FeB4RAG. However, in the specific context of RAG, alternative evaluation measures, as suggested by recent research in the field~\cite{gienapp2023evaluating}, might be more suitable. We note this is still an open line of research, but we believe our collection can support alternative evaluation paradigms.

%

\subsection{Expanding the Collection}

The design of FeB4RAG inherently accommodates future expansion and adaptation, aligning with the dynamic nature of advances in federated search and retrieval. We provide a comprehensive codebase that automates most of the processes involved in creating and updating the collection, thus facilitating potential expansions with minimal effort\footnote{\url{https://github.com/ielab/FeB4RAG}}.
Key expansion possibilities include:

\begin{itemize}[leftmargin=*]
	\item \textbf{Incorporation of New User Requests:} Our collection is structured to easily integrate new user requests, which can be derived from additional queries in the BEIR dataset. This capability allows for continuous enrichment and diversification of the collection.
	\item \textbf{Integration of New Dense Retrievers:} As newer and more specialized dense retriever models become available, they can replace or augment the current models used to determine the top-10 search results. This ensures that the collection remains at the forefront of retrieval technology. We also highlight that other search technologies, aside from dense retrievers, could be used to implement a search engine within our collection. The choice of limiting ghd search engines to be dense retrievers as opposed to other, more computationally expensive search technologies (e.g., zero-shot LLM rankers~\cite{wang2024zero,zhuang2023setwise,zhuang2023beyond}), was based purely on computational budget constraints.
	
	\item \textbf{Utilization of Advanced LLMs for Relevance Assessment:} Currently, we employ two high-performing LLMs to undertake relevance labelling. With the advent of more advanced LLMs, new models can be incorporated to enhance the quality and reliability of relevance labels, thereby continuously improving the collection's accuracy and usefulness.
	\item \textbf{Integration into Full RAG Pipelines:} The outcome from the result merging task can be integrated into a complete RAG pipeline. This integration facilitates direct evaluation through the quality of responses generated in a federated search-embedded RAG system. We demonstrate such an evaluation approach in the following section, highlighting the practical application of the collection in real-world scenarios, as well as the need for our collection.
\end{itemize}

%% file: sections/results.tex
 \begin{table}
	\scriptsize
	\centering
	\caption{Prompt for GPT-4 model to generate a response from user request and search results.}
	\label{table:generation_prompt}
	\resizebox{0.95\columnwidth}{!}{
		\begin{tabular}{p{2pt}|M{220pt}}
			\toprule
			& \textnormal{Prompt} \\
			\hline
			\rotatebox[origin=c]{90}{System Prompt} & You are a helpful assistant helping to answer user requests based on the provided search result. \linebreak Your responses should directly address the user's request and must be based on the information obtained from the provided search results. \linebreak  You are forbidden to create new information that is not supported by these results.  \linebreak You must attribute your response to the source from the search results by including citations, for example, [1]. \\ \midrule
			
			\rotatebox[origin=c]{90}{{\centering User Prompt}} & User Request: \linebreak \{request\} \linebreak Search Results:\linebreak \{search\_results\}\linebreak Response:\\
			\bottomrule
		\end{tabular}
		}

\end{table}

\section{Demonstrating the Need for FeB4RAG}
\label{sec:demo}
We motivated the creation of FeB4RAG with the need to evaluate federated search technology in the context of RAG pipelines, like those implemented by conversational agents. In this section we demonstrate the importance of having a tailored federated search approach within a RAG pipeline, compared to a naive approach to the federation of the resources. 


\subsection{RAG Systems with Federated Search}
We simulated two distinct federates search systems:
\begin{itemize}[leftmargin=*]
	\item \textbf{Naive-federated (\texttt{naive-fed})}: This system uses all search engines, with no resource selection taking place. Thus, an information request is routed to each of the search engines; then results from the individual search engines are merged using a round-robin strategy to form the aggregated result list, and the list is then culled to the top-k results. This in turn is provided as input to an LLM (via the prompt) to generate a text response to the information request.

	\item \textbf{Best-federated (\texttt{best-fed})}: This system performs resource selection to route the information request only to search engines that are relevant to the request. For this, for each information request, we rely on the search engine labels and select a search engine only if its graded precision is higher than zero. This simulates an optimal resource selection strategy. Then, search results are aggregated into a top-k ranking by considering only search results that are at least of minimal relevance (label score 1), and ranked in decreasing order of relevance. This simulates an optimal result merging strategy. This aggregated search engine result list in then provided as input to an LLM (via the prompt) to generate a text response to the information request.
	
\end{itemize}	
	
For both systems, we set  $k=16$, so that each resource contributes a single search result when operating the \texttt{naive-fed} system (round-robin). 
The search results, coupled with the information request, are fed into GPT-4\footnote{we use gpt-4-0125-preview and set temperature=0 for deterministic responses.}, following the prompt structure reported in Table~\ref{table:generation_prompt}, which is similar to those present in open-source RAG pipelines.

\subsection{Evaluation Methodology}
We conducted a pairwise evaluation of the responses generated by the LLM when using each of the two simulated federated systems above, utilizing a framework based on four criteria adapted from the work of \citet{gienapp2023evaluating}:
\begin{enumerate}[leftmargin=*]
	\item \textbf{Coverage}: This criterion evaluates the comprehensiveness of the information in addressing the user's query.
	\item \textbf{Consistency}: This criterion assesses whether the information aligns consistently with the cited sources.
	\item \textbf{Correctness}: This criterion measures the factual accuracy and reliability of the information provided.
	\item \textbf{Clarity}: This criterion gauges the ease of understanding the provided information from the user's perspective.
\end{enumerate}

\begin{figure}[t]
	\centering
	
	\includegraphics[width=\columnwidth]{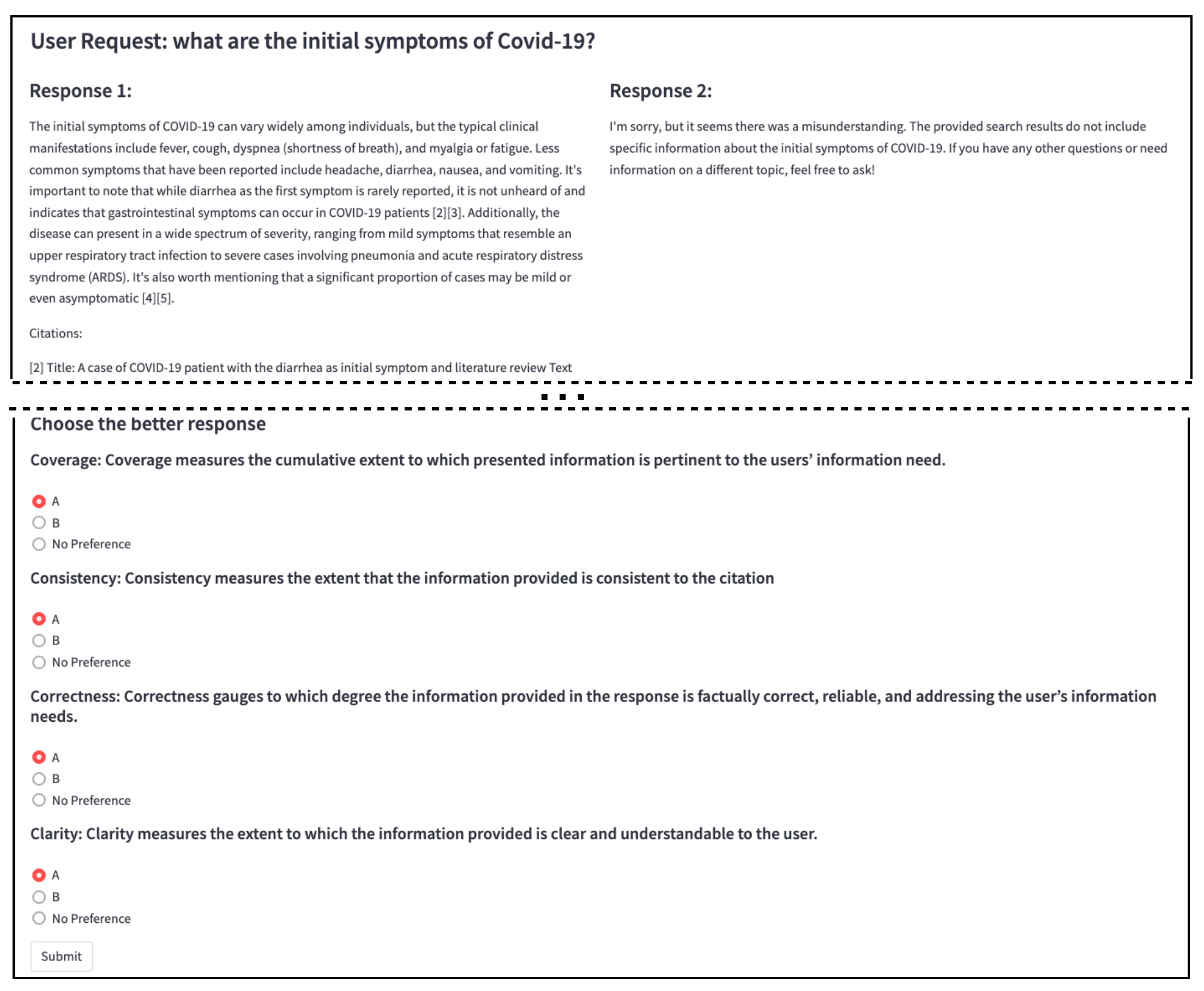}

	\caption{The evaluation interface we used for side-by-side preference comparison among RAG responses. Note that we cut the middle part of the page for space purposes -- the part of the image left out reported the remaining of response 1.}
	\label{fig:Example_judge}
\end{figure}

For the pairwise comparison evaluation process, we developed a web interface using Streamlit\footnote{\url{https://streamlit.io/}}; the interface is shown in Figure~\ref{fig:Example_judge}, and implements a side-by-side comparison~\cite{thomas2006evaluation}. To mitigate bias, the interface randomizes the presentation order of responses for each new information request. We then employed human assessors (the authors of this paper) to express preferences towards the responses. Assessors were instructed to select the best response among the two based on the specific evaluation criteria. If no clear preference is found, they were directed to select `No Preference`.

Then, a response is deemed better than the alternative by the counting of preferences towards each response with respect to the four criteria. Responses that receive an equal number of preferences are considered to be equivalent.  
To keep the evaluation workload within our hourly budget, we selectively sampled 80 requests from the FeB4RAG collection\footnote{However, we do include responses generated by the LLM for the whole set of requests as part of the published collection.}, choosing five requests from each BEIR dataset.

\begin{figure}[t]
	\centering

	\includegraphics[width=\textwidth]{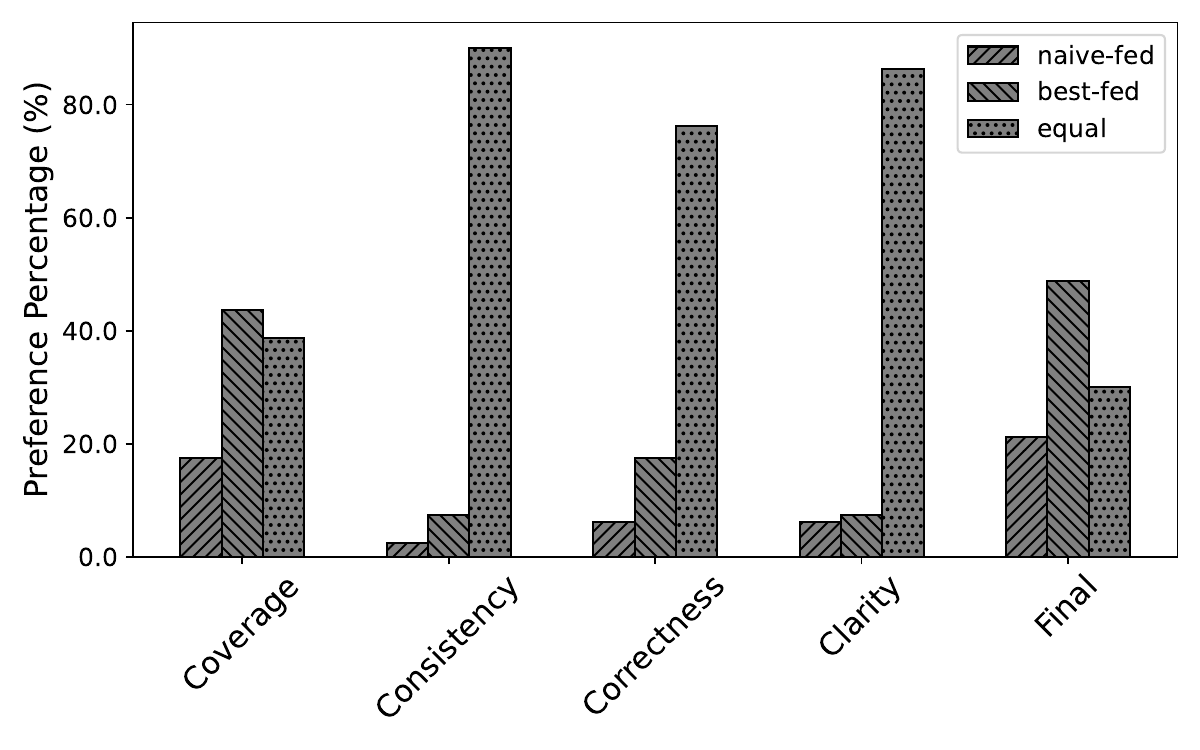}

	\caption{Pairwise evaluation of generated responses with respect to each evaluation criteria, comparing \textit{naive-fed} and \textit{best-fed} methodologies. \textit{Equal} indicates instances where no preference was observed between the two approaches. }
	\label{fig:winning_cases}
\end{figure}

\begin{figure}[t]
	\centering
	\includegraphics[width=\columnwidth]{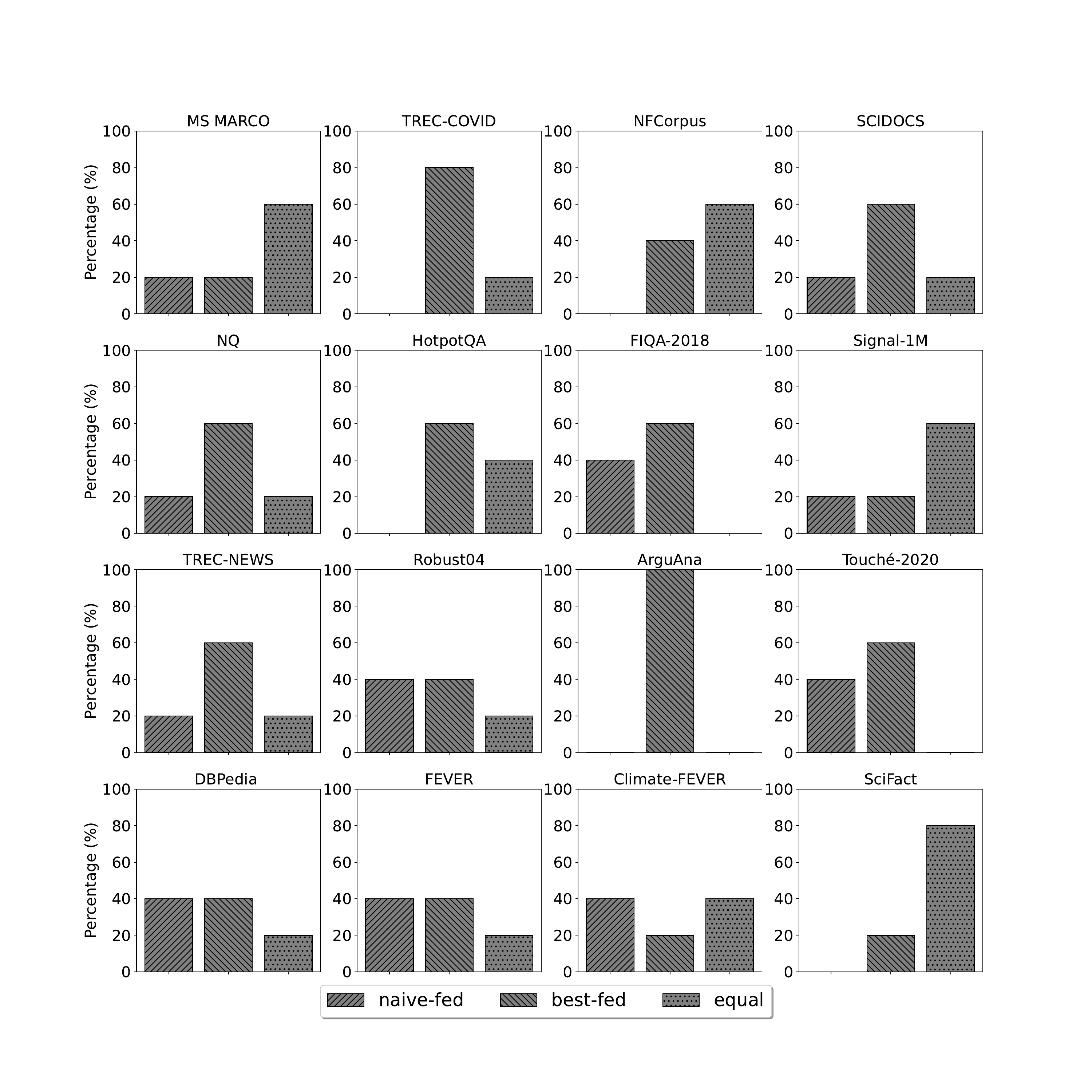}

	\caption{Preferences towards federated systems faceted according to the BEIR dataset from which the information request originated.}
	\label{fig:winning_individual}
\end{figure}

\subsection{Result Analysis}
Figure~\ref{fig:winning_cases} shows the breakdown of preferences across the four criteria, along with the final aggregated preferences. 
Responses generated using results from \texttt{best-fed} are preferred more often than those obtained using \texttt{naive-fed}: this occurs both at a criteria level (though differences for the clarity and consistency criteria are marginal) and when preferences are aggregated. Preferences for the coverage criteria exhibited the most differences across the two responses, with also correctness criteria exhibiting a smaller but comparatively sizeable difference.



Further insights are reported in Figure~\ref{fig:winning_individual}, where we facet the aggregated preferences with respect to the BEIR datasets from which the information requests originated. Recall that for the experiments in this section, we sampled from the set of 790 requests, with five requests for each dataset. The figure shows that preferences for responses generated by the RAG pipeline that rely on \texttt{best-fed} occur across all datasets; we note this is not the case for \texttt{naive-fed} (e.g. see, TREC-COVID and ArguAna). In addition, in most datasets \texttt{best-fed} receives a higher number of preferences compared to \texttt{naive-fed}, except for Climate-FEVER.

%% file: sections/conclusion.tex
\section{Conclusion}

This paper introduces FeB4RAG, a new test collection designed for evaluating federated search within RAG pipelines. The collection, in particular, provides a comprehensive framework to explore resource selection strategies and result merging techniques within RAG pipelines.
The key characteristics of our collection are:


\begin{itemize}[leftmargin=*]
	\item A set of 790 information requests, distinct from those in existing collections in that they more likely represent what users submit to RAG applications.
	
	\item Integration of 16 search engines built on top individual datasets of the BEIR benchmark, and based on state-of-the-art dense retrieval methods.
	\item Comprehensive relevance judgements obtained using  Large Language Models at both the search result and search engine levels, facilitating the evaluation of methods for federated search, as well as offering new avenues for evaluation.
	\item An expandable codebase that can be re-used and adapted to integrate more assessments, search engines and information requests, ensuring that FeB4RAG remains relevant for future RAG-related developments.
\end{itemize}

%